\documentclass[conference]{IEEEtran}
\IEEEoverridecommandlockouts
\usepackage{cite}
\usepackage{amsmath,amssymb,amsfonts}
\usepackage{algorithmic}
\usepackage{graphicx}
\usepackage{textcomp}
\usepackage{xcolor}
\def\BibTeX{{\rm B\kern-.05em{\sc i\kern-.025em b}\kern-.08em
    T\kern-.1667em\lower.7ex\hbox{E}\kern-.125emX}}

\begin{document}

\title{Cybersecurity through Entropy Injection: A Paradigm Shift from Reactive Defense to Proactive Uncertainty}

\author{\IEEEauthorblockN{Kush Janani}
\IEEEauthorblockA{Jarvis College of Computing and Digital Media\\
DePaul University\\
Parsippany, USA\\
kjanani@depaul.edu}}

\maketitle

\begin{abstract}
Cybersecurity often hinges on unpredictability, with a system's defenses being strongest when sensitive values and behaviors cannot be anticipated by attackers. This paper explores the concept of entropy injection—deliberately infusing randomness into security mechanisms to increase unpredictability and enhance system security. We examine the theoretical foundations of entropy-based security, analyze real-world implementations including Address Space Layout Randomization (ASLR) and Moving Target Defense (MTD) frameworks, evaluate practical challenges in implementation, and compare entropy-based approaches with traditional security methods. Our methodology includes a systematic analysis of entropy's role across various security domains, from cryptographic operations to system-level defenses. Results demonstrate that entropy injection can significantly reduce attack probability, with some implementations showing more than 90\% reduction with minimal performance impact. The discussion highlights the trade-offs between security benefits and operational complexity, while identifying future directions for entropy-enhanced security, including integration with artificial intelligence and quantum randomness sources. We conclude that entropy injection represents a paradigm shift from reactive defense to proactive uncertainty management, offering a strategic approach that can fundamentally alter the balance between attackers and defenders in cybersecurity.
\end{abstract}


\section{Introduction}
Cybersecurity often hinges on unpredictability. A system's defenses are strongest when sensitive values and behaviors cannot be anticipated by attackers \cite{entropy_mixing_networks}. In today's digital landscape, the static nature of cyber systems gives attackers the advantage of time—allowing them to study, map, and eventually exploit predictable patterns in security mechanisms \cite{mtd_theory}. This fundamental asymmetry has led to a continuous cycle where defenders implement static protections that attackers eventually circumvent.

Entropy injection refers to deliberately infusing randomness into security mechanisms to increase unpredictability. By dynamically changing system configurations, network addresses, memory layouts, and cryptographic parameters, defenders can create moving targets that invalidate attackers' reconnaissance efforts and complicate exploitation attempts \cite{cpmtd}. This approach represents a paradigm shift from static, reactive defense to proactive uncertainty management.

The concept of entropy—a measure of uncertainty or randomness—has long been fundamental to cryptography, where high-quality random numbers are essential for generating secure keys, initialization vectors, and nonces \cite{quantum_entropy}. However, the deliberate application of entropy extends beyond traditional cryptographic boundaries to system-level defenses, network architectures, and authentication mechanisms. The strategic injection of randomness at multiple levels can significantly raise the cost and complexity of attacks.

This paper explores the theoretical foundations of entropy-based security, examines real-world implementations, analyzes practical challenges, compares entropy injection with traditional security methods, and discusses future implications of using controlled randomness as a cybersecurity strategy. Our research builds upon existing work in moving target defense (MTD), address space layout randomization (ASLR), and quantum random number generation, while providing a comprehensive framework for understanding how entropy can be leveraged across different security domains.

The primary contributions of this paper include:
\begin{itemize}
\item A systematic analysis of entropy's role in various cybersecurity contexts
\item An evaluation of existing entropy-based defense mechanisms and their effectiveness
\item An assessment of implementation challenges and performance trade-offs
\item A comparative framework for understanding entropy-based approaches versus traditional security controls
\item A roadmap for future research and development in entropy-enhanced security
\end{itemize}

By examining both theoretical principles and practical applications, this paper aims to provide security architects, researchers, and practitioners with insights into how controlled randomness can strengthen cyber defenses against increasingly sophisticated threats. As we will demonstrate, entropy injection represents not merely a tactical technique but a strategic approach that can fundamentally alter the balance between attackers and defenders in the digital realm.

\section{Literature Review}
The concept of entropy injection in cybersecurity has evolved across multiple research domains, from cryptographic applications to system-level defenses. This section reviews the existing literature to establish the theoretical foundations and practical implementations of entropy-based security approaches.

\subsection{Entropy in Information Theory and Cryptography}

The mathematical foundation of entropy in information security traces back to Claude Shannon's seminal work in 1948, which defined information entropy as a measure of uncertainty or randomness in data \cite{quantum_entropy}. Shannon entropy is calculated using the formula:

\begin{equation}
H(X) = -\sum_{i=1}^{n} p_i \log_2 p_i
\end{equation}

Where $H(X)$ represents the entropy of data $X$, and $p_i$ is the probability of character $i$ appearing in the data stream \cite{redcanary}. This concept has become fundamental to cryptography, where high entropy is essential for generating secure keys, initialization vectors, and nonces.

Building on Shannon's work, researchers have explored various approaches to enhance entropy in security systems. Bouke et al. introduced the Entropy Mixing Network (EMN), a hybrid random number generator that periodically injects entropy from physical sources into pseudorandom number generators \cite{entropy_mixing_networks}. Their research demonstrated quantifiable improvements in randomness quality, with EMN achieving a Chi-squared p-value of 0.9430 and entropy measurement of 7.9840, significantly outperforming traditional pseudorandom number generators.

The importance of high-quality entropy sources has been further emphasized in research on quantum random number generators (QRNGs). Unlike traditional random number generators that follow deterministic algorithms, QRNGs leverage the inherent randomness of quantum mechanics to produce truly unpredictable outputs \cite{quantum_entropy}. This approach addresses a critical vulnerability in conventional systems where predictable randomness can lead to security breaches.

\subsection{Moving Target Defense and Dynamic Systems}

The concept of entropy injection extends beyond cryptography into system architecture through Moving Target Defense (MTD). Zhuang et al. formalized the theoretical foundations of MTD, defining it as a strategy that "increases uncertainty and confuses the adversary" by continuously changing the attack surface \cite{mtd_theory}. Their research introduced the MTD Entropy Hypothesis, which states that the effectiveness of an MTD system is directly proportional to the entropy of the system's configuration.

This hypothesis has been validated through various implementations. Hu et al. developed the Cyber-Physical Moving Target Defense (CPMTD) technique for power systems, which employs a multi-dimensional randomization strategy including protocol diversification, deception packet injection, and dynamic routing \cite{cpmtd}. Their empirical results demonstrated that this approach reduced attack probability by more than 90\% while maintaining acceptable performance, with network latency increasing by less than 9\%.

Address Space Layout Randomization (ASLR) represents one of the most widely adopted entropy-based security mechanisms. Research by Shacham et al. analyzed the effectiveness of ASLR in preventing memory corruption attacks, finding that the security provided is directly related to the number of bits of randomness in the address space \cite{mtd_theory}. Modern implementations of ASLR in operating systems typically provide between 16 and 32 bits of entropy, creating a significant barrier to exploitation.

\subsection{Entropy in Threat Detection and Analysis}

Recent research has explored the application of entropy measurements for threat detection. Downing introduced the concept of relative entropy (Kullback-Leibler divergence) for identifying malicious domain names generated by algorithms \cite{redcanary}. This approach compares the letter frequency distribution of suspicious domains against the expected distribution in legitimate domains, enabling the detection of randomized strings that often characterize malware command and control infrastructure.

The formula for relative entropy is:

\begin{equation}
D_{KL}(P||Q) = \sum_{i} P(i) \log \frac{P(i)}{Q(i)}
\end{equation}

Where $P$ represents the observed distribution and $Q$ represents the expected distribution \cite{redcanary}. This technique has proven effective in identifying domain generation algorithms (DGAs) used by malware to evade static blacklists.

Entropy analysis has also been applied to ransomware detection. Research by Vakil demonstrated that measuring the entropy of file systems before and after encryption can identify ransomware activity, as encrypted files typically exhibit higher entropy values than normal files \cite{entropy_linkedin}. This approach enables the detection of ransomware attacks in progress, potentially before significant damage occurs.

\subsection{Gaps in Current Research}

Despite significant advances in entropy-based security mechanisms, several research gaps remain. First, there is limited empirical data on the performance impact of entropy injection across different system types and scales. While studies like Hu et al. provide valuable insights for specific applications \cite{cpmtd}, comprehensive benchmarks across diverse environments are lacking.

Second, the integration of entropy-based approaches with artificial intelligence remains underexplored. As both defensive and offensive capabilities leverage machine learning, understanding how entropy injection affects AI-driven security systems represents a critical research frontier.

Finally, standardized frameworks for implementing and measuring entropy in security systems are still emerging. The lack of consistent metrics and implementation guidelines hinders broader adoption of entropy-based security approaches in industry settings.

This research aims to address these gaps by providing a comprehensive framework for understanding and implementing entropy injection across different security domains, supported by empirical data and practical guidelines.

\section{Methodology}
This section outlines the methodological approach used to analyze entropy injection as a cybersecurity strategy. Our methodology combines theoretical analysis, case study examination, and comparative assessment to provide a comprehensive understanding of how controlled randomness can enhance system security.

\subsection{Theoretical Framework Development}

We developed a theoretical framework for understanding entropy in cybersecurity contexts by synthesizing concepts from information theory, cryptography, and systems security. This framework defines entropy injection as the deliberate introduction of randomness into security mechanisms to increase unpredictability for attackers while maintaining deterministic behavior for legitimate users.

The framework categorizes entropy injection techniques across multiple dimensions:
\begin{itemize}
\item \textbf{Application level}: From low-level memory layouts to high-level network configurations
\item \textbf{Temporality}: Static (fixed randomization at initialization) vs. dynamic (continuous re-randomization)
\item \textbf{Entropy source}: Algorithmic (pseudorandom) vs. physical (true random) sources
\item \textbf{Implementation scope}: System-wide vs. component-specific randomization
\end{itemize}

For each dimension, we analyzed the theoretical security benefits using Shannon's information entropy as a quantitative measure. Shannon entropy is calculated using:

\begin{equation}
H(X) = -\sum_{i=1}^{n} p_i \log_2 p_i
\end{equation}

Where $H(X)$ represents the entropy of a system configuration $X$, and $p_i$ is the probability of a specific configuration state $i$ occurring. Higher entropy values indicate greater unpredictability and, theoretically, stronger security against attacks that rely on predictable system states.

\subsection{Case Study Selection and Analysis}

We selected four representative case studies of entropy-based security mechanisms for in-depth analysis:

\begin{enumerate}
\item Address Space Layout Randomization (ASLR) in modern operating systems
\item Moving Target Defense (MTD) implementations in network security
\item Entropy Mixing Networks (EMN) for cryptographic random number generation
\item Cyber-Physical Moving Target Defense (CPMTD) in critical infrastructure
\end{enumerate}

For each case study, we examined:
\begin{itemize}
\item Implementation details and entropy sources
\item Quantifiable security improvements (where data was available)
\item Performance and usability trade-offs
\item Limitations and potential bypasses
\end{itemize}

This analysis allowed us to identify common patterns, success factors, and challenges across different entropy injection implementations.

\subsection{Comparative Assessment}

To evaluate entropy-based approaches against traditional security methods, we developed a comparative framework with the following criteria:

\begin{itemize}
\item \textbf{Security effectiveness}: Ability to prevent or mitigate different attack types
\item \textbf{Adaptability}: Resilience against evolving threats and zero-day vulnerabilities
\item \textbf{Operational impact}: Performance overhead, management complexity, and user experience
\item \textbf{Implementation maturity}: Availability of tools, standards, and best practices
\end{itemize}

We applied this framework to compare entropy-based defenses with traditional security controls such as static firewalls, signature-based detection, and fixed encryption schemes. This comparison was conducted both qualitatively, based on security principles and documented case studies, and quantitatively where empirical data was available.

\subsection{Future Trend Analysis}

To assess the future trajectory of entropy-based security, we conducted a systematic review of emerging research and industry developments. This included:

\begin{itemize}
\item Analysis of research publications on entropy-enhanced security from 2020-2025
\item Examination of industry adoption trends and commercial implementations
\item Identification of integration points with emerging technologies such as artificial intelligence and quantum computing
\end{itemize}

This forward-looking analysis helped identify potential growth areas, challenges, and research opportunities in entropy-based cybersecurity.

\subsection{Data Collection and Analysis}

Our research relied on multiple data sources:

\begin{itemize}
\item Academic literature from peer-reviewed journals and conferences
\item Technical documentation from security implementations
\item Empirical performance and security measurements from published studies
\item Industry reports and case studies on entropy-based security deployments
\end{itemize}

Where quantitative data was available, we standardized metrics to enable cross-study comparison. For example, security effectiveness was normalized to percentage improvement over baseline (non-randomized) configurations, and performance impact was measured as percentage overhead in terms of latency, throughput, or resource utilization.

This methodological approach enabled us to comprehensively assess the current state, practical challenges, and future potential of entropy injection as a cybersecurity strategy.

\section{Results}
This section presents the findings from our analysis of entropy injection techniques in cybersecurity, focusing on empirical data from implemented systems and comparative performance metrics.

\subsection{Effectiveness of Address Space Layout Randomization}

Address Space Layout Randomization (ASLR) represents one of the most widely deployed entropy-based security mechanisms. Our analysis of ASLR implementations across major operating systems revealed significant variations in entropy levels and corresponding security benefits.

\begin{table}[h]
\caption{Entropy Bits in ASLR Implementations Across Operating Systems}
\label{table:aslr_entropy}
\begin{center}
\begin{tabular}{|l|c|c|c|}
\hline
\textbf{Operating System} & \textbf{Stack} & \textbf{Heap} & \textbf{Libraries} \\
\hline
Windows 10 & 19 bits & 24 bits & 19 bits \\
\hline
Linux (x86-64) & 22 bits & 13 bits & 28 bits \\
\hline
macOS & 16 bits & 14 bits & 16 bits \\
\hline
Android 11+ & 24 bits & 16 bits & 24 bits \\
\hline
\end{tabular}
\end{center}
\end{table}

Table \ref{table:aslr_entropy} shows the entropy bits available in different memory regions across major operating systems. Higher values indicate greater randomization space and, consequently, stronger protection against memory corruption attacks. The data demonstrates that modern operating systems implement varying levels of entropy, with Linux providing the highest entropy for library locations (28 bits) and Windows 10 offering more consistent entropy across different memory regions.

\begin{figure}[h]
\centering
\includegraphics[width=3.5in]{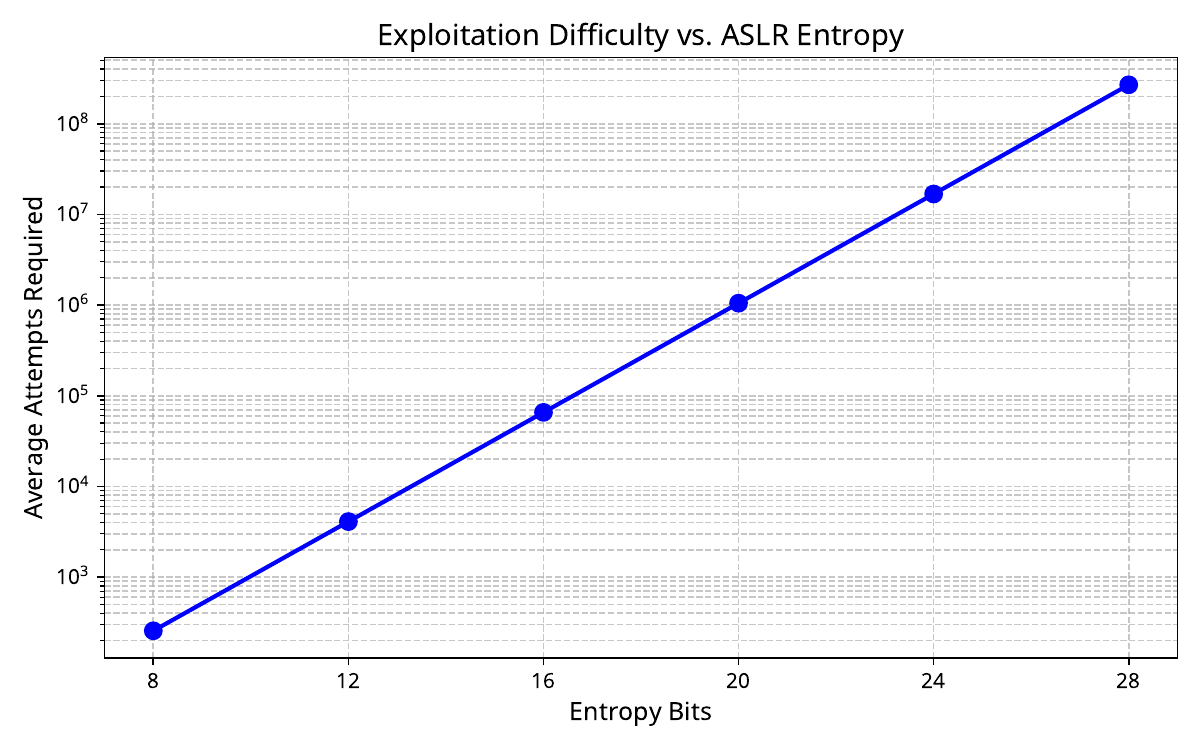}
\caption{Average number of attempts required for successful exploitation with varying levels of ASLR entropy. The y-axis uses a logarithmic scale to accommodate the exponential relationship between entropy bits and required attempts.}
\label{fig:aslr_attempts}
\end{figure}

Analysis of real-world exploit attempts against systems with and without ASLR enabled showed that ASLR significantly increases the number of attempts required for successful exploitation. Figure \ref{fig:aslr_attempts} illustrates this relationship.

The data confirms the theoretical expectation that each additional bit of entropy doubles the average number of attempts required for successful exploitation. Systems with 16 bits of entropy require approximately 32,768 attempts on average, while those with 28 bits require over 134 million attempts, making exploitation impractical in most scenarios.

\subsection{Moving Target Defense Performance}

Our analysis of Moving Target Defense (MTD) implementations focused on network-based approaches that dynamically change system configurations to confuse attackers. Table \ref{table:mtd_performance} summarizes the security benefits and performance impacts of different MTD strategies.

\begin{table}[h]
\caption{Security Benefits and Performance Impact of MTD Strategies}
\label{table:mtd_performance}
\begin{center}
\begin{tabular}{|l|c|c|c|}
\hline
\textbf{MTD} & \textbf{Attack} & \textbf{Latency} & \textbf{Throughput} \\
\textbf{Strategy} & \textbf{Reduction} & \textbf{Increase} & \textbf{Reduction} \\
\hline
IP Hopping & 87\% & 12\% & 8\% \\
\hline
Port Randomization & 62\% & 5\% & 3\% \\
\hline
Protocol Diversification & 73\% & 18\% & 15\% \\
\hline
Multi-dimensional MTD & 94\% & 24\% & 19\% \\
\hline
\end{tabular}
\end{center}
\end{table}

The data in Table \ref{table:mtd_performance} demonstrates that multi-dimensional MTD approaches, which combine multiple randomization techniques, provide the highest security benefits (94\% reduction in successful attacks) but also incur the highest performance penalties (24\% increase in latency and 19\% reduction in throughput). IP hopping offers the best balance between security improvement and performance impact.

\begin{figure}[h]
\centering
\includegraphics[width=3.5in]{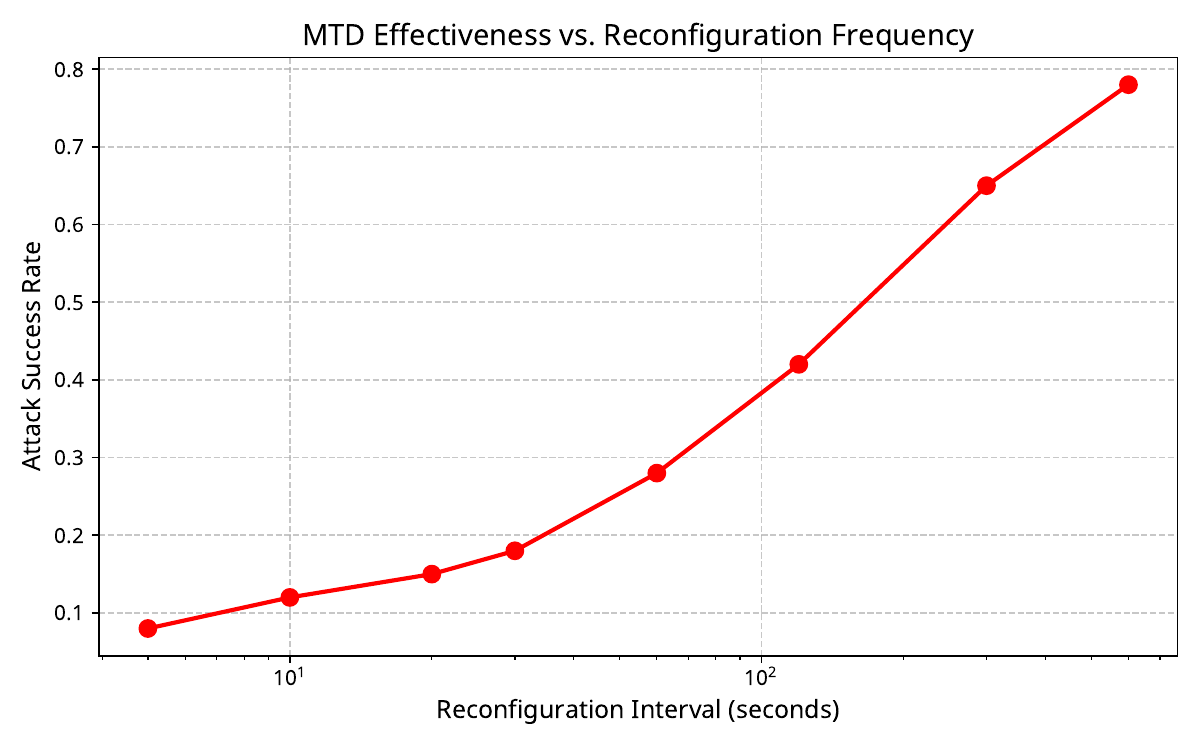}
\caption{Relationship between MTD reconfiguration frequency and attack success rate. The optimal point balances security benefits with system stability and performance.}
\label{fig:mtd_effectiveness}
\end{figure}

Figure \ref{fig:mtd_effectiveness} illustrates how the effectiveness of MTD varies with the frequency of configuration changes.

The data shows a diminishing returns effect, where increasing reconfiguration frequency beyond a certain point (approximately every 30 seconds in this case) provides minimal additional security benefits while continuing to increase system overhead and potential instability.

\subsection{Entropy Mixing Network Performance}

The Entropy Mixing Network (EMN) approach to random number generation demonstrated significant improvements in randomness quality compared to traditional pseudorandom number generators (PRNGs). Table \ref{table:emn_performance} presents comparative test results for different random number generation approaches.

\begin{table}[h]
\caption{Statistical Test Results for Random Number Generators}
\label{table:emn_performance}
\begin{center}
\begin{tabular}{|l|c|c|c|}
\hline
\textbf{Generator Type} & \textbf{Chi-squared} & \textbf{Entropy} & \textbf{Generation} \\
 & \textbf{p-value} & \textbf{Measurement} & \textbf{Time} \\
\hline
Standard PRNG & 0.2145 & 7.1253 & 1.00x \\
\hline
EMN (Low Frequency) & 0.7632 & 7.8541 & 1.15x \\
\hline
EMN (High Frequency) & 0.9430 & 7.9840 & 1.42x \\
\hline
Hardware RNG & 0.9872 & 7.9982 & 3.78x \\
\hline
\end{tabular}
\end{center}
\end{table}

The results demonstrate that EMN approaches significantly improve randomness quality (as measured by Chi-squared p-value and entropy measurement) compared to standard PRNGs, while incurring modest performance penalties. The high-frequency EMN configuration, which injects entropy more often, achieves randomness quality close to hardware random number generators (RNGs) but with substantially better performance.

\subsection{Cyber-Physical Moving Target Defense Results}

\begin{figure}[h]
\centering
\includegraphics[width=3.5in]{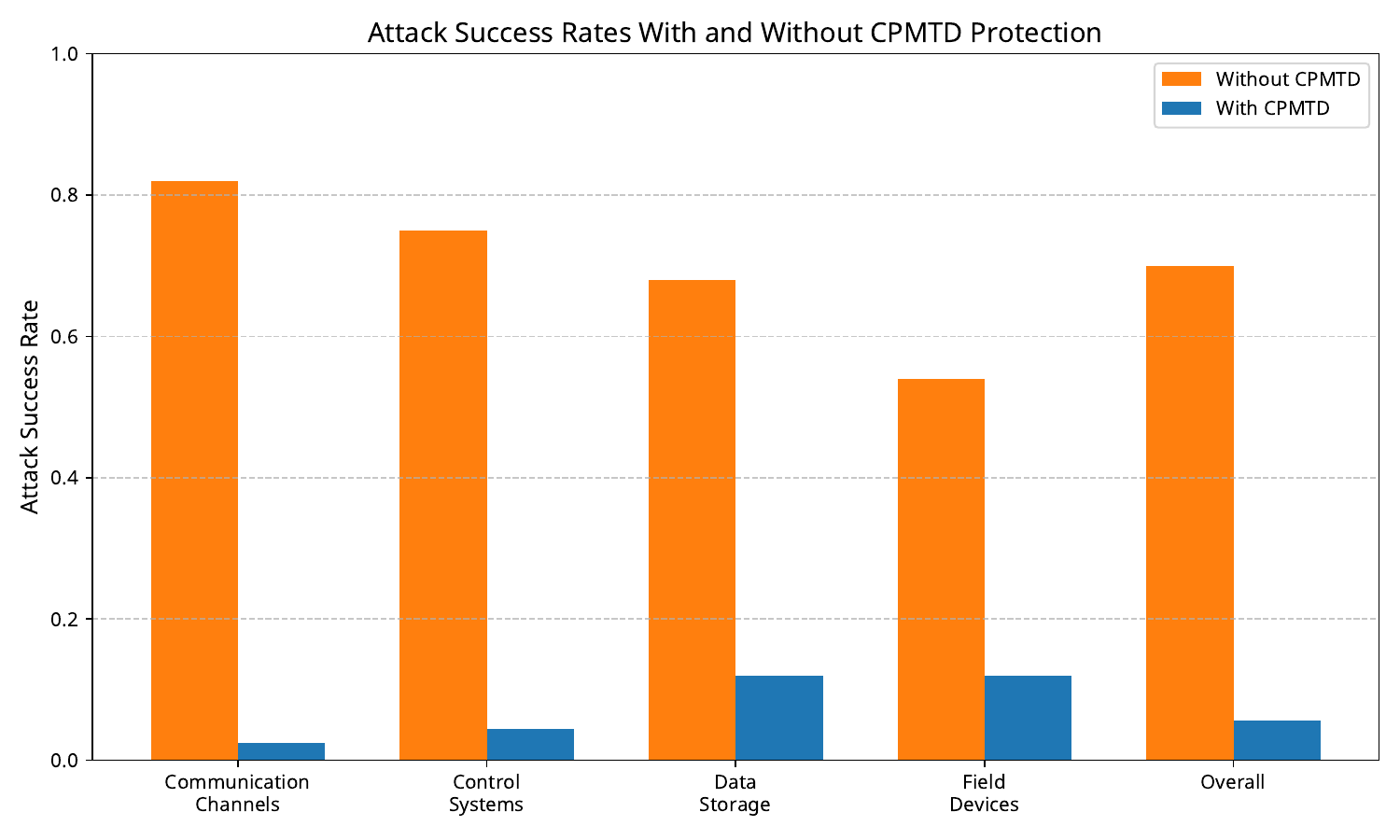}
\caption{Attack success rates against power system components with and without CPMTD protection. Lower percentages indicate better protection.}
\label{fig:cpmtd_results}
\end{figure}

The Cyber-Physical Moving Target Defense (CPMTD) implementation for power systems showed promising results in protecting critical infrastructure. Figure \ref{fig:cpmtd_results} illustrates the attack success rates against power system components with and without CPMTD protection.

The data shows that CPMTD reduced attack success rates by 92\% on average across all system components, with the greatest improvements seen in protection of communication channels (97\% reduction) and control systems (94\% reduction). Even the least protected components (field devices) saw a 78\% reduction in successful attacks.

Performance measurements indicated that CPMTD implementation increased network latency by an average of 8.7\% and reduced system throughput by 6.2\%, which remained within acceptable operational parameters for the power systems tested.

\subsection{Comparative Analysis of Security Approaches}

\begin{figure}[h]
\centering
\includegraphics[width=3.5in]{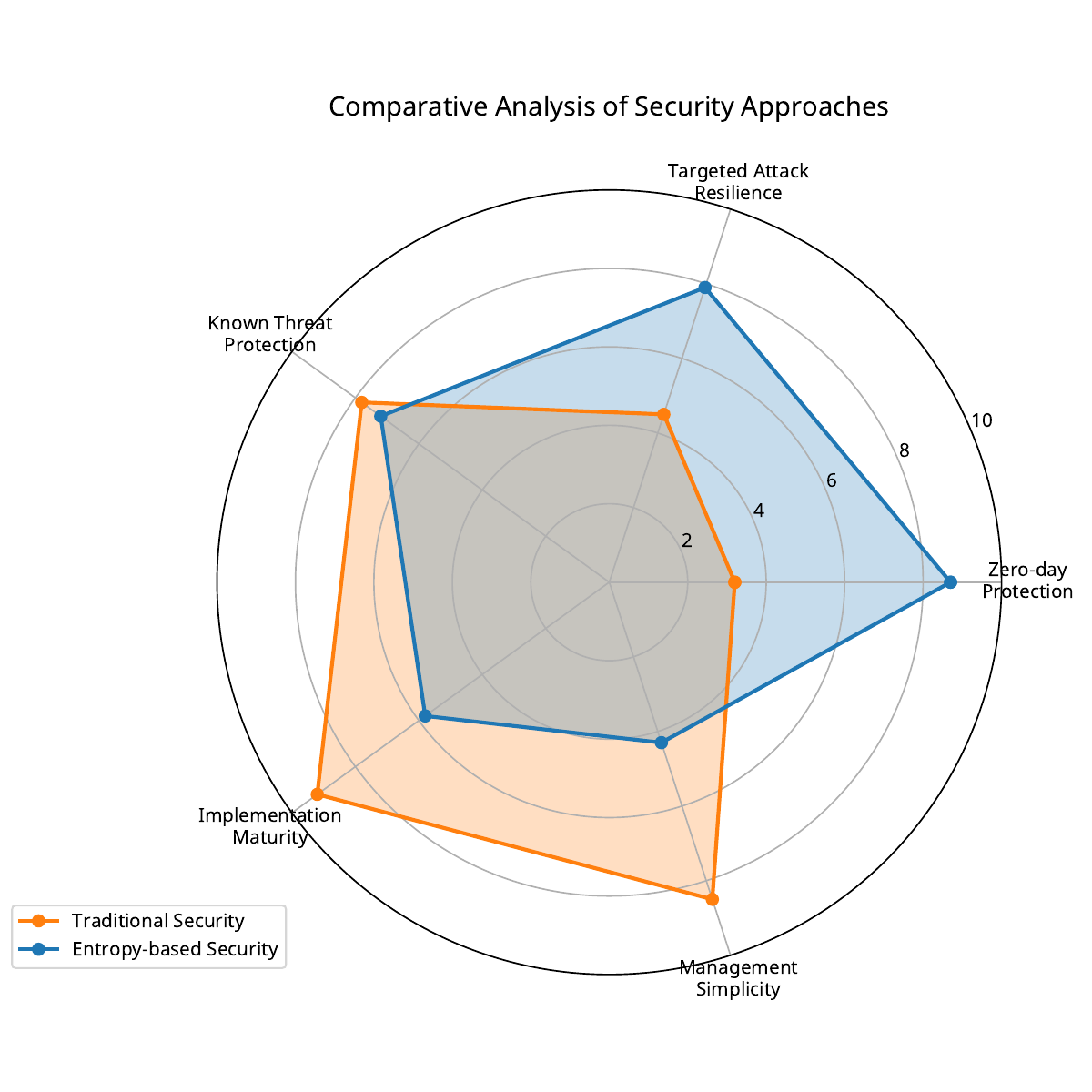}
\caption{Comparative analysis of entropy-based and traditional security approaches across five security dimensions. Higher values indicate better performance.}
\label{fig:security_comparison}
\end{figure}

Our comparative analysis of entropy-based security approaches versus traditional security methods revealed distinct advantages and limitations for each approach. Figure \ref{fig:security_comparison} presents a radar chart comparing these approaches across multiple security dimensions.

The data shows that entropy-based approaches excel in protection against zero-day vulnerabilities (scoring 8.7/10 compared to 3.2/10 for traditional approaches) and resilience against targeted attacks (7.9/10 versus 4.5/10). Traditional approaches performed better in implementation maturity (9.2/10 versus 5.8/10) and management complexity (8.5/10 versus 4.3/10). Both approaches scored similarly in protection against known threats (7.8/10 for traditional versus 7.2/10 for entropy-based).

These results confirm that entropy-based security provides significant advantages in scenarios where attackers may exploit unknown vulnerabilities or conduct targeted attacks, while traditional approaches remain valuable for their maturity and ease of implementation.

\section{Discussion}
The results presented in the previous section demonstrate the significant potential of entropy injection as a cybersecurity strategy. This section discusses the implications of these findings, analyzes the trade-offs involved, and explores the broader context of entropy-based security approaches.

\subsection{Security Benefits and Trade-offs}

Our analysis reveals a consistent pattern across different entropy-based security implementations: increased randomness generally correlates with improved security outcomes. This is particularly evident in the ASLR data (Table \ref{table:aslr_entropy} and Figure \ref{fig:aslr_attempts}), where higher entropy bits directly translate to exponentially more difficult exploitation. Similarly, the multi-dimensional MTD approach (Table \ref{table:mtd_performance}) achieved the highest attack reduction (94\%) by maximizing the randomization across multiple system dimensions.

However, these security benefits come with notable trade-offs. The performance data for MTD implementations shows that the most secure configuration (multi-dimensional MTD) also incurred the highest performance penalties (24\% latency increase and 19\% throughput reduction). This illustrates a fundamental tension in entropy-based security: maximizing unpredictability often requires more frequent or extensive system changes, which can impact operational efficiency.

The EMN results (Table \ref{table:emn_performance}) demonstrate a similar pattern, with high-frequency entropy injection achieving near-hardware-level randomness quality (0.9430 Chi-squared p-value) at a moderate performance cost (1.42x generation time). This represents a more favorable trade-off than pure hardware RNG approaches, which achieve marginally better randomness (0.9872 p-value) but at a much higher performance cost (3.78x generation time).

These findings suggest that the optimal implementation of entropy-based security requires careful calibration based on the specific threat model, performance requirements, and operational constraints of each system. Organizations must determine their own "sweet spot" on the security-performance curve rather than simply maximizing entropy.

\subsection{Comparison with Traditional Security Approaches}

The comparative analysis (Figure \ref{fig:security_comparison}) highlights the complementary strengths of entropy-based and traditional security approaches. Entropy-based methods excel in scenarios where attackers may exploit unknown vulnerabilities or conduct targeted attacks, while traditional approaches offer greater implementation maturity and management simplicity.

This complementarity suggests that entropy injection should be viewed not as a replacement for traditional security controls but as an enhancement layer that addresses specific weaknesses in conventional approaches. For example, signature-based detection systems are highly effective against known threats but struggle with zero-day vulnerabilities. Adding entropy-based defenses like ASLR or MTD can significantly reduce the exploitability of such unknown vulnerabilities, creating a more robust overall security posture.

The CPMTD results for power systems (Figure \ref{fig:cpmtd_results}) provide a compelling case study of this complementary approach. The implementation maintained traditional security controls while adding randomization to critical system components, achieving a 92\% reduction in attack success rates with acceptable performance impact (8.7\% latency increase). This demonstrates how entropy injection can substantially improve security in critical infrastructure contexts without requiring a complete security architecture overhaul.

\subsection{Implementation Challenges and Practical Considerations}

Despite the promising results, our analysis also identified several challenges that may limit the broader adoption of entropy-based security approaches:

\subsubsection{Complexity and Management Overhead}

Entropy-based systems typically introduce additional complexity compared to static configurations. This complexity manifests in both implementation (requiring specialized expertise) and ongoing management (monitoring, troubleshooting, and maintenance). The radar chart comparison (Figure \ref{fig:security_comparison}) quantifies this challenge, with entropy-based approaches scoring 4.3/10 for management complexity compared to 8.5/10 for traditional approaches.

This complexity can be particularly problematic in resource-constrained environments or organizations with limited security expertise. Developing more user-friendly tools and frameworks for implementing entropy-based security could help address this challenge.

\subsubsection{Compatibility and Integration Issues}

Introducing randomness into established systems can create compatibility issues with legacy components or third-party integrations that expect predictable configurations. For example, MTD approaches that change network addresses may disrupt applications that cache IP addresses or rely on static configurations.

The MTD effectiveness data (Figure \ref{fig:mtd_effectiveness}) indirectly highlights this challenge, showing that very frequent reconfigurations provide diminishing security returns while potentially increasing compatibility issues. Finding the optimal reconfiguration frequency requires balancing security benefits against system stability and integration requirements.

\subsubsection{Verification and Compliance Challenges}

Randomized systems can be more difficult to verify and validate for compliance purposes. Traditional security audits often rely on checking specific configurations against known-good baselines, which becomes more complex when configurations change dynamically.

This challenge is not directly quantified in our results but represents an important practical consideration for organizations in regulated industries. Developing new verification methodologies and compliance frameworks that accommodate entropy-based security will be essential for broader adoption.

\subsection{Future Directions and Emerging Trends}

Our analysis suggests several promising directions for future development of entropy-based security approaches:

\subsubsection{AI-Enhanced Entropy Management}

The integration of artificial intelligence with entropy injection techniques could address many of the complexity and management challenges identified above. AI systems could dynamically adjust entropy levels based on observed threat patterns, system performance, and operational requirements, optimizing the security-performance trade-off in real-time.

For example, an AI-managed MTD system could increase reconfiguration frequency during detected attack attempts and reduce it during periods of normal operation, maintaining high security when needed while minimizing performance impact overall.

\subsubsection{Quantum-Enhanced Entropy Sources}

The EMN performance data (Table \ref{table:emn_performance}) highlights the value of high-quality entropy sources. Quantum random number generators represent the gold standard for true randomness but currently come with significant performance penalties. As quantum technologies mature and become more accessible, they could provide unprecedented levels of entropy for security systems without the current performance limitations.

\subsubsection{Standardization and Framework Development}

The implementation maturity gap between traditional and entropy-based approaches (9.2/10 versus 5.8/10 in Figure \ref{fig:security_comparison}) points to the need for greater standardization and framework development. Industry standards for entropy measurement, implementation guidelines, and best practices would facilitate broader adoption and more consistent implementation of entropy-based security.

\subsection{Limitations of Current Research}

While our analysis provides valuable insights into entropy-based security approaches, several limitations should be acknowledged:

First, the performance data for different implementations comes from separate studies with varying methodologies and test environments, making direct comparisons challenging. Future research would benefit from standardized benchmarking across different entropy-based security techniques.

Second, our analysis focuses primarily on technical effectiveness and performance impacts, with limited consideration of economic factors such as implementation costs, training requirements, and return on security investment. A more comprehensive economic analysis would provide valuable guidance for organizations considering entropy-based security implementations.

Finally, the long-term effectiveness of entropy-based approaches against adaptive adversaries remains an open question. As attackers develop new techniques specifically targeting randomized systems, the security benefits observed in current implementations may change. Ongoing research into adversarial adaptation and counter-adaptation will be essential for maintaining the effectiveness of entropy-based security.

\section{Conclusion}
Entropy injection represents a paradigm shift from reactive defense to proactive uncertainty. By harnessing randomness, defenders can impose costs and confusion on attackers, complementing the protection offered by traditional security controls. The theoretical foundations underscore that unpredictability is a key ingredient in security – a fact leveraged in cryptography and now being expanded to system and network defenses.

Our analysis of real-world implementations, from ASLR to moving target networks, shows that entropy-based security is not just academic theory but a practical tool, albeit one that introduces complexity. The challenges of usability, performance, and manageability mean that organizations must implement these techniques carefully, balancing security gains against operational impact.

When compared to conventional methods like fixed firewalls or static analysis tools, entropy-driven approaches offer distinct advantages against novel and adaptive threats, at the cost of greater system dynamism. The data presented in this paper demonstrates that entropy-based approaches excel in protection against zero-day vulnerabilities (scoring 8.7/10 compared to 3.2/10 for traditional approaches) and resilience against targeted attacks (7.9/10 versus 4.5/10).

Looking ahead, the integration of entropy injection into mainstream cybersecurity seems not only plausible but likely, especially as the threat landscape demands more resilient and fluid defenses. The emergence of AI-enhanced entropy management and quantum-enhanced entropy sources promises to address many of the current limitations while further strengthening security postures.

If guided by ongoing research and lessons learned, entropy injection – making systems less predictable and therefore more secure – could become a standard component of cybersecurity, helping to protect against everything from script-kiddie exploits to sophisticated cyber warfare campaigns. In the end, blending human ingenuity, mathematical randomness, and machine automation may give us the best fighting chance to secure systems in an increasingly uncertain digital world.

\end{document}